\documentclass[12pt]{article}
\usepackage{a4wide}
\usepackage{floatflt}
\usepackage{amssymb}
\usepackage{graphicx}
\usepackage{amsmath}
\usepackage[center]{subfigure}

\newcommand{\ba}{\begin{eqnarray}}
\newcommand{\ea}{\end{eqnarray}}
\newcommand{\be}{\begin{equation}}
\newcommand{\ee}{\end{equation}}
\begin{document}
\begin{titlepage}
\begin{flushright}
SI-HEP-2004-06\\
SFB/CPP-04-29
\end{flushright}
\vfill
\begin{center}
{\Large\bf Kaon Distribution Amplitude from QCD Sum Rules }\\[2cm]
{\large\bf A.~Khodjamirian, Th.~Mannel and M.~Melcher}\\[0.5cm]
{\it Theoretische Physik 1, Fachbereich Physik,
Universit\"at Siegen,\\ D-57068 Siegen, Germany }\\
\end{center}
\vfill
\begin{abstract}
We present a new calculation of the first Gegenbauer moment $a_1^K$
of the kaon light-cone distribution amplitude.
This moment is determined by the difference
between the average momenta
of strange and nonstrange valence quarks in the kaon.
To calculate $a_1^K$, QCD sum rule for the
diagonal correlation function
of local and nonlocal axial-vector currents is used.
Contributions of condensates up
to dimension six are taken into account,
including $O(\alpha_s)$-corrections to the
quark-condensate term.
We obtain $a_1^K=0.05\pm 0.02$,
differing by the sign and magnitude from
the recent sum-rule estimate from the nondiagonal correlation function
of pseudoscalar and axial-vector currents.
We argue that the nondiagonal sum rule is numerically not reliable.
Furthermore, an independent
indication for a positive $a_1^K$ is given,
based on the matching of two different
light-cone sum rules for the $K\to\pi$ form factor.
With the new interval of $a_1^K$,
we update our previous numerical
predictions for
SU(3)-violating effects in  $B_{(s)}\to K$ form
factors and charmless \(B\) decays.
\end{abstract}
\vfill

\end{titlepage}

\newcommand{\DS}[1]{/\!\!\!#1}

\section{Introduction}

Light-cone distribution amplitudes (DA's) of hadrons
are universal long-distance objects involved
in many QCD approaches to exclusive
hadronic processes  with a large energy- or mass-scale.
The variety of these processes (from form factors at
large momentum transfers to  heavy-meson exclusive decays),
and the diversity of the approaches
(from effective theories to QCD sum rules) makes the determination
of DA's  an important task.
In this paper we will mainly concentrate on
the twist-2 DA of the kaon, defined by the standard expression
\begin{equation}
\langle K^-(q)|\bar{s}(0)\gamma_\mu\gamma_5\left[0,z\right]u(z)|0
\rangle_{z^2\to 0}
  = -i q_\mu f_K\int_0^1 du~e^{i\bar{u}q\cdot z}\varphi_K(u,\mu)\,,
\label{eq-phiK}
\end{equation}
where $\bar{u}=1-u$, $[0,z]$ is the path-ordered gauge-factor (Wilson line)
and $\mu$ is
the normalization scale determined by the interval $z^2$ near the light-cone.

The twist-2 DA $\varphi_K(u)$ is usually expanded in
Gegenbauer polynomials
\begin{equation}
  \varphi_K(u,\mu)=6u\bar{u}\left(1+\sum_{n=1}^\infty a_n^K(\mu)C_n^{3/2}(u-\bar{u})\right),
\label{eq-moments}
\end{equation}
with the multiplicatively renormalizable
coefficients $a_n^K(\mu)$ (Gegenbauer moments). Taken at some low scale 
$\mu \sim 1$ GeV,
the moments $a_n^K(1 \mbox{GeV})$ encode the long-distance dynamics.
The anomalous dimension of $a_n^K$ grows  with $n$. Hence, in many applications
of DA's  where a high normalization scale is involved, the higher moments
are suppressed, and only the lowest moments $a_{1,2}^K$ are retained.

As opposed to the pion case,
where the odd moments $a_{1,3,..}^\pi$ vanish in the isospin symmetry limit,
the first moment $a_{1}^K$ is expected to be as important as $a_2^K$.
The nonzero value of $a_1^K$ reveals a flavour-SU(3)
violation effect of $O(m_{s}-m_{u,d})$. In physical terms,
$a_1^K$ is proportional to  the difference
between the longitudinal momenta of the strange and nonstrange quark
in the two-particle Fock component of the kaon. In  our definition (\ref{eq-phiK})
these fractions are $x_s=u$ and $x_{\bar{u}}=\bar{u}$,
respectively, so that
\be
\langle x_s-x_{\bar u}\rangle_{K^-}=
\int\limits_0^1 du ~\varphi_K(u)(u-\bar{u})= \frac35a_1^K\,.
\label{a1u}
\ee
Hence, knowing or at least constraining  $a_1(1 \mbox{GeV})$ 
is indispensable for predicting SU(3)-violation effects
within any QCD approach that employs  DA's. The accurate knowledge of these 
effects is particularly important in  $B$ decays to pions and kaons, in 
the context of CP-violation and CKM-matrix studies.

As originally suggested in \cite{CZ}, the few first Gegenbauer 
moments of DA's  can be calculated employing QCD sum rules 
\cite{SVZ} based on the
local operator-product expansion (OPE) of a dedicated correlation
function of two quark currents.
To estimate $a_1^K$, two different correlation
functions of kaon-interpolating quark currents
have been considered in \cite{CZ}: the {\em diagonal} one,
with two axial-vector currents
and the {\em nondiagonal} one,
with one pseudoscalar and one axial-vector current.
In what follows we call also the sum rules obtained from the 
respective correlation functions diagonal and nondiagonal. 
In both cases there is an axial-vector current 
$\bar{s}\gamma_\mu\gamma_5\!\stackrel{\leftrightarrow}{D}_\alpha\!u$ 
containing a covariant derivative (where $\stackrel{\leftrightarrow}{D}=
\stackrel{\rightarrow}{D}-\stackrel{\leftarrow}{D}$). The matrix element of 
this current entering the kaon term in the sum rule is proportional to $a_1^K$:
\be
\langle K^-(q)|\bar{s}\gamma_\mu\gamma_5i\!\stackrel{\leftrightarrow}{D}_\alpha\!u|0\rangle
= -iq_\mu q_\alpha f_K \frac35 a_1^K\,.
\label{eq-matrel}
\ee
This equation is easily obtained by expanding (\ref{eq-phiK})
near $z=0$ and retaining the leading-twist term. 
In \cite{CZ} only the leading-order contributions to OPE
were taken into account and both sum rules  predicted a positive value
\be
[a_1^K(1 \mbox{GeV})]^{CZ}\sim 0.17\,,
\label{CZa1}
\ee
yielding
$\langle x_s-x_{\bar{u}}\rangle_{K^-}\simeq 0.10$.
The positive sign of the SU(3)-breaking asymmetry is in accordance
with an intuitive expectation for the kaon constituents:
the heavier strange quark (antiquark)
has a larger momentum fraction than the
lighter nonstrange antiquark (quark). Although in contrast 
to intuition, also a negative sign of $a_1^K$ is possible,
since the strange quark is not sufficiently heavy with respect to the
typical hadronic scales, e.g. $m_s(1~\mbox{GeV})\sim\Lambda_{QCD}$.

Recently, the nondiagonal QCD sum rule for $a_1^K$,
together with the sum rules for the moments of $K^*$-meson DA's,
have been reconsidered and updated in \cite{BallB}. The
$O(\alpha_s)$ radiative corrections to the perturbative and quark-condensate terms
in this sum rule have also been calculated.
Importantly, the authors of \cite{BallB}
have traced a sign error in the leading perturbative term
of the nondiagonal sum rule for $a_1^K$ written down in \cite{CZ}.
Correcting the sign and adding radiative corrections produce 
a drastic change in the hierarchy of terms
in this sum rule. As a result, the value of $a^K_1$
has changed to  \cite{BallB}
\be
[a_1^K(1 \mbox{GeV})]^{BB} = -0.18\pm 0.09.
\label{BBnumber}
\ee
We have used the latter estimate in \cite{KMM},
in obtaining predictions on SU(3)-violation
in exclusive $B$ decays and form factors.

As we already noted in \cite{KMM}, in order to get an independent
estimate of $a_1^K$, one has to return to the diagonal
correlation function and update the corresponding sum rule with
the same $O(\alpha_s)$ accuracy.
In this paper this task is to a large extent fulfilled.
We undertake a new calculation
of \(a_1^K\) (and in parallel of $a_{2}^K$)
from the diagonal correlation function keeping all terms
up to dimension six in OPE and including the
\(O(\alpha_s)\)-correction to the
quark-condensate contribution.
Quite surprisingly, our result
\be
a_1^K(\mu\sim 1 \mbox{GeV})= 0.05\pm 0.02\,,
\label{a1number}
\ee
substantially differs
from the prediction (\ref{BBnumber}) of the nondiagonal sum rule \cite{BallB}.

The paper is organized as follows.
The details of the sum rule derivation are
presented in Section 2, and the numerical analysis is
performed in Section 3.
In order to examine the contradiction between our result
and the one obtained in \cite{BallB}, in Section 4
we have a closer look at the nondiagonal sum rule. We find
that it is rather difficult to obtain
a reliable numerical estimate of $a_1^K$ from that sum rule.
In section 5 we also give an additional independent argument
in favour of positive $a_1^K$, employing light-cone sum rules
for the spacelike $K\to\pi$ transition form factor.
Furthermore, with the new estimate  (\ref{a1number}),
in section 6 we update the numerical estimates for SU(3)-violating
effects in $B$ decays obtained in \cite{KMM}. Our conclusions are presented in
section 7.

\section{QCD sum rule for $a_n^K$}

The way to derive QCD sum rules for the Gegenbauer coefficients
is thoroughly explained in \cite{CZ} and also in \cite{BallB}. In the latter paper,
instead of operators with covariant derivatives,
an elegant device of nonlocal operators is employed, which we also use
here. The underlying (diagonal) correlation function
is chosen as
\begin{equation}
  \Pi(q^2,q\cdot z)=i\int d^4x ~e^{i q\cdot x}\langle 0|T\left\{\bar{u}(x)\DS{z}\gamma_5 s(x),
  \bar{s}(0)\DS{z}\gamma_5\left[0,z\right]u(z)\right\}|0\rangle,
\label{eq-corr}
\end{equation}
where \(z^2=0\) and
we use an auxiliary nonlocal operator instead of a current with a fixed
number of covariant derivatives.
By inserting a complete set of hadronic states with $K$-meson quantum numbers
in (\ref{eq-corr}) we obtain
\ba
&&\Pi^{hadr}(q^2,q\cdot z)= (q\cdot z)^2\frac{f_K^2
\int_0^1 du~e^{i\bar{u}q\cdot z} \varphi_K(u)}{m_K^2-q^2}
\nonumber\\
&&+\sum\limits_{K_h}\frac{\langle 0|\bar{u}\DS{z}\gamma_5 s |K_h\rangle
\langle K_h| \bar{s}(0)\DS{z}\gamma_5\left[0,z\right]u(z)|0\rangle}{m_{K_h}^2-q^2}\,,
\label{hadr}
\ea
where the ground-state contribution of the kaon
is shown explicitly and the sum takes into account the excited resonances
and continuum states with the kaon quantum numbers. In the above we
used (\ref{eq-phiK}) and
the definition of the decay constant
$\langle 0 | \bar{u}\gamma_\alpha \gamma_5 s|K^-(q)\rangle=if_Kq_\alpha$.
The raw sum rule is obtained by equating (\ref{hadr})
to the result of OPE for $\Pi(q^2,q\cdot z)$ in terms of perturbative
and condensate contributions.
Below, the OPE result will be cast in the form
\begin{equation}
\Pi^{OPE}(q^2,q\cdot z)=(q\cdot z)^2\int_0^1 du~e^{i\bar{u} q\cdot z}\pi(u,q^2),
\label{eq-ope}
\end{equation}
where \(\pi(u,q^2)\) may also contain delta function of $u$
and its derivatives. In addition we need the dispersion
relation in $q^2$ for this function:
\be
\pi(u,q^2)=\frac{1}{\pi}\int\limits_{0}^\infty ds\frac{\mbox{Im}_s\pi(u,s)}{s-q^2-i\epsilon}\,,
\label{eq-disp}
\ee
where subtraction terms are not essential, since they will
vanish after Borel transformation.
Employing quark-hadron duality,
the sum over higher state contributions in (\ref{hadr}) is then
approximated by the integral (\ref{eq-disp}) where the lower limit
is replaced by a certain effective threshold $s_0^K$.
Subtracting this integral (the {\em continuum contribution}) from both parts of the
equation $\Pi^{hadr}=\Pi^{OPE}$ and performing Borel transformation
we obtain a generic sum rule for the DA:
\be
f_K^2\int\limits_0^1 du\, e^{i\bar{u}q\cdot z} \varphi_K(u)e^{-m_K^2/M^2}
=\int_0^1 du~e^{i\bar{u} q\cdot z}\frac{1}{\pi}\int\limits_{0}^{s_0^K}ds~
e^{-s/M^2}\mbox{Im}_s\pi(u,s)\,.
\label{sumrule1}
\ee
In order to project out the $n$-th moment and to obtain the
sum rule for $a_n^K$ one has to replace
$$
e^{i\bar{u} q\cdot z}\to C_n^{3/2}(u-\bar{u})\,,
\label{eq-replace}
$$
in both parts of this equation. The result is:
\begin{equation}
  a_n^K=\frac{2(2n+3)}{3(n+1)(n+2)}
\left(\frac{e^{m_K^2/M^2}}{f_K^2}\right)\frac{1}{\pi}\int_{0}^{s_0^K}ds~
  e^{-s/M^2} \int_0^1 du~C_n^{3/2}(u-\bar{u})\mbox{Im}_s \pi(u,s),
\label{eq-sumrule}
\end{equation}
where the $n$-dependent factor comes from the normalization of Gegenbauer polynomials.
At $n=0$, $a_0^K=1$ and (\ref{eq-sumrule}) turns into the original SVZ sum rule
for $f_K^2$ \cite{SVZ} (see also \cite{KMM} where the SU(3)-violation
in this sum rule is investigated). In fact,
only the few first moments of (\ref{eq-sumrule}) are useful,
practically $n\leq 2$, because
as already realized in \cite{CZ} the higher-dimensional condensate contributions
rapidly grow with $n$ and one cannot rely on
local OPE \footnote{We will not discuss
an interesting possibility to introduce nonlocal condensates \cite{MikhailRad}
which is a different approach.}. A few comments are in order concerning
the sum rule (\ref{eq-sumrule}). First,
the threshold parameter $s_0^K$ generally depends on $n$. Usually
 $(s_0^K)_n$ is fitted together with $a_n$ to achieve the maximal stability in the
relevant region of Borel parameter where both higher-dimensional condensates
and the continuum contribution (the duality estimate for higher states)
are reasonably small, say,
not exceeding 30\% both. Second, the scale at which $a_n$ is estimated
from the QCD sum rules is of the order of $M$, the characteristic virtuality of the
correlation function. Since $M$ covers an interval around 1 GeV it is just the
scale we need.

The QCD calculation of the master function $\pi(u,q^2)$ in a form
of OPE is straightforward and we 
only give a few accompanying comments while  explicitly representing the
results. We expand the correlation function up to dimension 6,  with the 
highest-dimension term in OPE coming from the four-quark condensate.
The small parameter in the correlation function is the ratio
$m_s/M$, therefore
the Wilson coefficients are also expanded in the strange quark mass.
The $u$- and $d$-quark masses are put to zero.
Generally, we neglect all terms where the power of $m_s/M$ added
to the dimension of the condensate exceeds 6. Furthermore, we use
the Fock-Schwinger gauge for the vacuum gluon field,
so that the gauge factor $[0,z]=1$.

The relevant contributions to the correlation function
are listed below:

\textbf{Perturbative term}: As the calculation
of this contribution given in the leading order 
by the loop diagram is not very complicated,
we kept the full \(m_s\) dependence. We get:
\begin{equation}
   \pi^{loop}(u,q^2)=-\frac{3 u\bar{u}}{2 \pi^2}\log\left(\frac{m_s^2-uq^2}{\mu^2}\right)\,,
\label{piloop}
\end{equation}
where $\mu$ is an arbitrary renormalization scale and
the polynomial terms in $q^2$ are not shown. They,
together with the $\mu$-dependent part  vanish after Borel
transformation.
At present state, $O(\alpha_s)$ corrections to the diagonal
correlation function are available only in
the massless limit \cite{Gorsky,MikhailRad,BallB}. 
For \(n=0\) and at \(O(m_s^0)\), the \(\alpha_s\)-correction
to (\ref{piloop}) is less than 10\%. 
In the sum rule for $a_1^K$, the leading-order perturbative  
contribution given by the convolution of (\ref{piloop})
with $C_1^{3/2}(2u-1)$  starts at $O(m_s^2)$ and is numerically suppressed 
with respect to the quark-condensate term.
For our calculation of $a_1$, we therefore assume that the $O(\alpha_s m_s^2)$ 
correction to the perturbative loop can be neglected. 
This conjecture can be verified in future with a direct calculation.

\textbf{Quark condensate}: At tree level, we get
\begin{equation}
  \pi^{\langle\bar{q}q\rangle}(u,q^2)=
  \frac{m_s \langle\bar{s}s\rangle}{(q^2)^2}\left[\left(1-\frac{1}{3}\frac{m_s^2}{q^2}\right)\delta(u)
  -\frac{1}{3}\frac{m_s^2}{q^2} \delta'(u)\right]\,,
\label{eq-cond}
\end{equation}
where a generic notation $\langle \bar{q}q\rangle \equiv \langle 0|\bar{q}q|0\rangle$,
$q=u,s$, is used for the quark-condensate density.
As  (\ref{eq-cond}) is the dominant contribution, we also calculated the \(O(\alpha_s)\)-correction. This time we employed the Feynman gauge for the perturbative gluons,
yielding additional contributions from the Wilson line.
Since our main goal is the sum rule for $a_1^K$,
we show here only the $n=1$ moment of the correlation function: 
\ba
 \int\limits _0^1 du \,C_1^{3/2}(u-\bar{u}) \pi^{\alpha_s\langle\bar{q}q\rangle}(u,q^2)&=&
  \frac{\alpha_s C_F}{\pi}\frac{m_s\langle\bar{s}s\rangle}{(q^2)^2}
  \left[2\left(\Delta-\log\left(-\frac{q^2}{\mu^2}\right)\right)+\frac{31}{3}\right]
\nonumber
\\
&-&\frac{\alpha_sC_F}{\pi}\frac{m_s\langle\bar{u}u\rangle}{(q^2)^2},
\label{alphasqq}
\ea
where $\Delta=2/(4-D)-\gamma_E+\log 4\pi$.
The corresponding expressions for $n=0,2,3$ are given in appendix \ref{App-Condensate-2}.
Here, we neglected higher orders of \(m_s\).
The combination \(m_s\langle\bar{s}s\rangle\) has zero anomalous
dimension and the ultraviolet divergence is absorbed by
the renormalization of \(a_1^K\). Note that the contribution of the $u$-quark
condensate absent in the leading order appears in $O(\alpha_s)$.\\

\textbf{Gluon condensate}: In calculating this contribution
described by the quark-loop diagrams with an emission of two vacuum gluons,
an additional term from the quark condensate
has to be taken into account (see, e.g. \cite{RRY} for details).
We obtain:
\begin{align}
\pi^{\langle G^2\rangle}(u,q^2)
=\frac{\langle G^2 \rangle}{12}&
 \Biggl[\frac{m_s^2\,u(2 u\,q^2+(1-3u)m_s^2)}{(m_s^2-u\,q^2)^4}\nonumber\\
       &+\frac{\delta(u)}{(q^2)^2}\left(1-\frac{m_s^2}{3\,q^2}\right)
       +\frac{\delta(\bar{u})}{2(m_s^2-q^2)^2}
       -\frac{\delta'(u)\,m_s^2}{3\,(q^2)^3}\Biggr],
\end{align}
where \(\langle G^2\rangle=\left<0\left|\frac{\alpha_s}{\pi}G_{\mu\nu}G^{\mu\nu}\right|0\right>\) is the gluon-condensate density.
Note that the terms in \(O(m_s^0)\) cancel after convolution with 
odd Gegenbauer polynomials. In particular, for the \(a_1^K\) sum rule 
the contribution
of the gluon condensate term is proportional to \(m_s^2\),
in accordance with the SU(3)-symmetry limit and with the chirality structure 
of the correlation function.

\textbf{Quark-gluon condensate.}
Here two effects contribute, as usual: the vacuum-gluon emission
from the virtual quark line and the local expansion of the light-quark fields
(up to three derivatives). The result reads:
\begin{equation}
  \pi^{\langle\bar{s}G s\rangle}(u,q^2)=
  \frac{1}{3}m_s\langle\bar{s}G s\rangle\delta'(u)\frac{1}{(q^2)^3}\,,
\end{equation}
where the notation \(\langle \bar{s}Gs\rangle
=\left<0\left|g_s\bar{s}\,\sigma_{\mu \nu}\frac{\lambda^a}{2}G^{a\mu\nu}s\right|0\right>\) for the quark-gluon condensate density is used.
The contribution of the quark-gluon condensate with  $u$ quarks vanishes
in the $m_u=0$ limit.

\textbf{Four-quark condensate}:
For this contribution, we employ the usual vacuum saturation ansatz \cite{SVZ}
factorizing all four-quark operators to the product of two quark condensates. 
The result reads:
\begin{align}
  \pi^{\langle\bar{q}q\rangle^2}(u,q^2)=&
  -\frac{32\pi\alpha_s}{81\,(q^2)^3}\Bigl(
    \langle\bar{s}s\rangle^2\left[\delta(u)+\delta'(u)\right]
   +\langle\bar{u}u\rangle^2\left[\delta(\bar{u})+\delta'(\bar{u})\right]\nonumber\\
   &+\frac{9}{2}\langle\bar{s}s\rangle\langle\bar{u}u\rangle
     \left[\delta(u)+\delta(\bar{u})\right]
   \Bigr),
\end{align}
where the term proportional to 
\(\langle\bar{u}u\rangle\langle\bar{s}s\rangle\) vanishes for odd moments.

Summing up all contributions to $\pi(u,q^2)$ listed above, 
we use the obtained expression for the OPE of the correlation function 
to calculate various moments from (\ref{eq-sumrule}). 
In particular, the desired sum rule for $a_1^K$ reads:
\ba
a_1^{K}=\frac{e^{m_K^2/M^2}}{f_K^2}\Bigg\{
\frac{5m_s^4}{4\pi^2}\int_{m_s^2}^{s_0^K}ds~e^{-\frac{s}{M^2}}\frac{(m_s^2-s)^2}{s^4}
-\frac{5}{3}\frac{m_s\langle\bar{s}s\rangle}{M^2}
  \left(1+\frac{1}{2}\frac{m_s^2}{M^2}\right)
\nonumber
\\
+\frac{5\alpha_sC_F}{9\pi}\frac{m_s}{M^2}\Bigg[
-\langle\bar{u}u\rangle
+2\langle\bar{s}s\rangle\left(
    \frac{25}{6}+\gamma_E-e^{-s_0^K/M^2}\frac{M^2}{s_0^K}
    -\log\left(\frac{M^2}{\mu^2}\right)
    -\mbox{Ei}\left(-\frac{s_0^K}{M^2}
\right)\right)
\Bigg]
\nonumber
\\
+\frac{5}{18\pi}\frac{m_s^2}{M^4}\langle G^2\rangle
\Bigg[
  \gamma_E-\frac{1}{4}-\mbox{Ei}\left(-\frac{s_0^K}{M^2}\right)+\log\left(\frac{m_s^2}{M^2}\right)
  +e^{-s_0^K/M^2}\left(\frac{M^4}{s_0^{K 2}}-\frac{M^2}{s_0^K}\right)
\Bigg]
\nonumber\\
+\frac{5}{9}\frac{m_s\langle\bar{s}G s\rangle}{M^4}
-\frac{80\pi}{81}
\frac{\alpha_s}{M^4}\left(\langle\bar{s}s\rangle^2-\langle\bar{u}u\rangle^2\right)
\Bigg\},
\label{eq-a1sumrule}
\ea
where the perturbative term starts from $O(m_s^2)$:
\be
\nonumber
\frac{5m_s^4}{4\pi^2}\int_{m_s^2}^{s_0^K}ds~e^{-\frac{s}{M^2}}\frac{(m_s^2-s)^2}{s^4}= \frac{5m_s^2}{12\pi^2}+O(m_s^4)\,,
\ee
and $\mbox{Ei}(x)=-\int_{-x}^{\infty} dt~ e^{-t}/t$.
The  $O(m_s^2)$ loop contribution
and the leading-order quark and quark-gluon condensate contributions
to this sum rule
have already been derived in \cite{CZ}, the other
terms are new. Our result for $\pi(u,q^2)$ allows to obtain also the even
moments in (\ref{eq-sumrule}). We have compared them with
the corresponding expressions in \cite{BallB}
(where the diagonal sum rule for $a_n^K$ is only applicable
for even $n$), and found agreement up to the terms of higher order in \(m_s\)
which are not present in \cite{BallB}.

\section{Numerical analysis}
To analyze the sum rules numerically,
we use the following intervals of
the relevant input parameters: \(m_s(1~\mbox{GeV})=130\pm 20~\mbox{MeV}\),
\(\langle\bar{q}q\rangle( 1\mbox{GeV})=
-(240\pm 10~\mbox{MeV})^3\),
\(\langle\bar{s}s\rangle=(0.8\pm 0.3)\langle\bar{q}q\rangle\),
\(\langle\bar{q}Gq\rangle(\mu)=(0.8\pm 0.2~\mbox{GeV}^2)\langle\bar{q}q\rangle\)(1 GeV)
(with a negligible scale-dependence) and
\(\langle(\alpha_s/\pi)G^2\rangle=0.012 \pm 0.006~\mbox{ GeV}^4\). Finally, we choose the
renormalization scale to be \( \mu=M\) and adopt $\alpha_s(M)$ with
$\bar{\Lambda}_{QCD}^{(n_f=3)}=320$ MeV.

In order to check the input for the diagonal sum rule, we first
analyze the \(n=0\) moment of (\ref{eq-sumrule}), which
yields the well-established SVZ sum rule for \(f_K\) \cite{SVZ}.
We are now in a position to improve the accuracy of this sum rule
adding the higher orders
in \(m_s\) and \(O(\alpha_s\langle\bar{q}q\rangle)\) corrections. To this end,
we fitted the threshold parameter \(s_0^K\) in order to achieve the
maximal stability
(weakest dependence on \(M^2\)) within the optimal interval of the Borel parameter
(the Borel window). The latter is the region of \(M^2\) in which the
OPE converges safely and excited states are suppressed. We found that for \(0.5~\mbox{GeV}^2<M^2<1.0~\mbox{GeV}^2\)
both the dimension-six four-quark condensate term
and the subtracted continuum contribution  in the sum rule are less than
\(30\%\). In that region,
the maximal stability is achieved for \(s_0^K=1.05~\mbox{GeV}^2\). As a result, we get from
the zeroth moment of (\ref{eq-sumrule}) $f_K=(0.92 \pm 0.02)f_K^{exp}$.
This agreement adds more confidence in the validity of the
lowest moments of the diagonal sum rule.

\begin{figure}
  \centering
  \subfigure[]{
    \centering\shortstack{\includegraphics[width=0.46\textwidth]{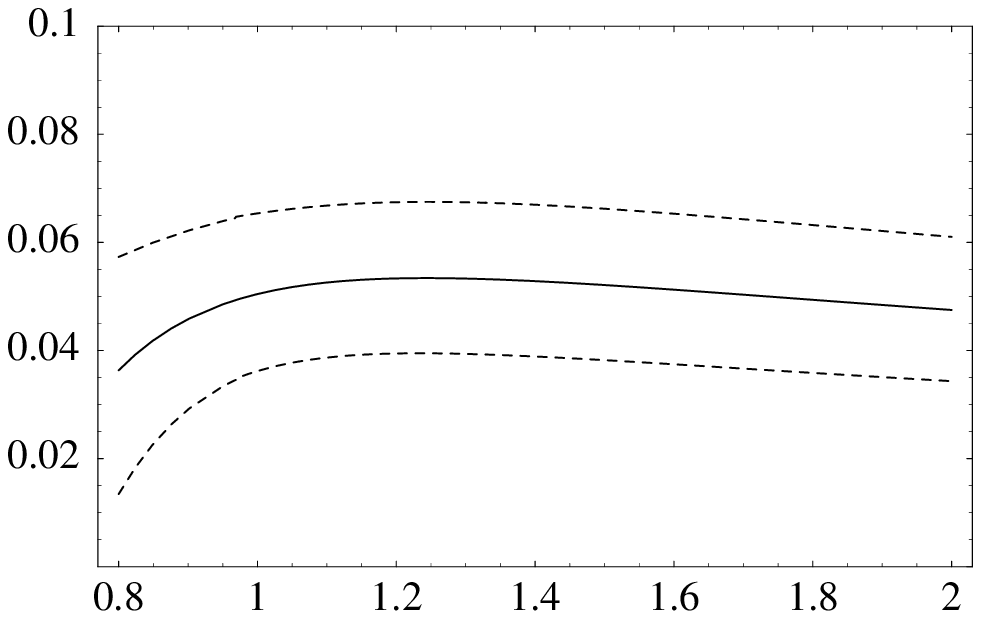}\\
             \small\(M^2/GeV^2\)} \label{fig-a1a}
    }%
  \subfigure[]{
    \centering\shortstack{\includegraphics[width=0.46\textwidth]{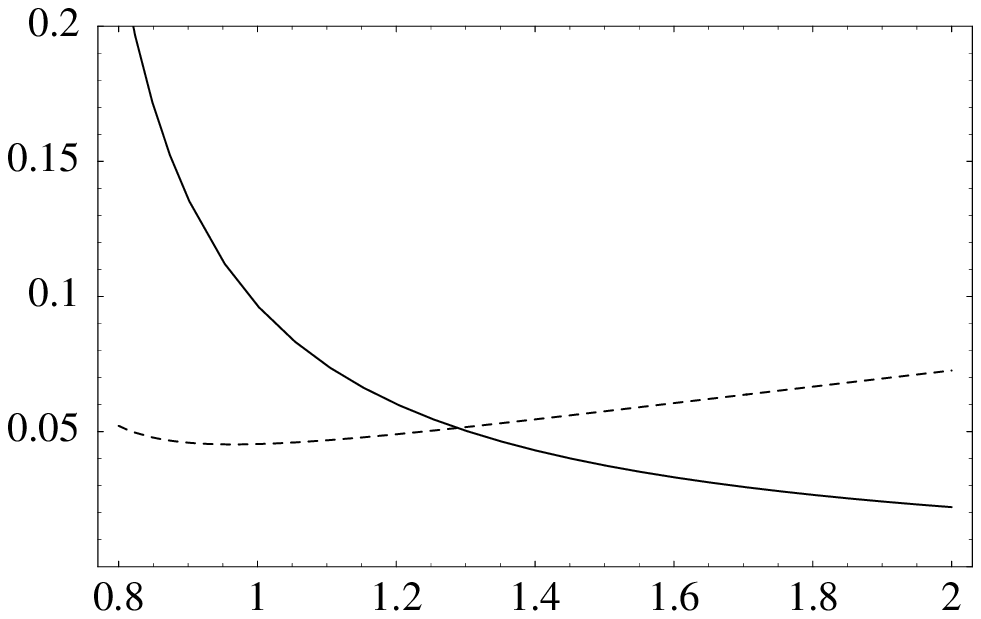}\\
             \small\(M^2/GeV^2\)} \label{fig-a1b}
}%
  \caption{{\it (a) The Gegenbauer moment \(a_1^K\)
evaluated from the diagonal sum rule
(\ref{eq-a1sumrule}) as a function of the Borel parameter $M^2$
(the dotted lines indicate the uncertainties induced by the input parameters);
(b) contributions from the dimension-six condensates (solid line)
and from the continuum (dotted line), divided
by the r.h.s. of (\ref{eq-a1sumrule}).}}
  \label{fig-a1-all}
\end{figure}

For \(a_1\) the numerical prediction of the sum rule
(\ref{eq-a1sumrule}) is shown in Fig. \ref{fig-a1a}.
Here one has to move the Borel window to \(M^2>0.8~\mbox{GeV}^2\), in order
to keep the OPE convergent. In the same
region, the \(\alpha_s\)-correction to the condensate term is \(<50\%\)
of the zeroth order, so that one can trust  the perturbative expansion.
On the other hand, the continuum contribution is less than \(10\%\)
for \(M^2\) up to \(1.5~\mbox{GeV}\).
The contributions from the continuum and from the dimension-six term of the
OPE are shown in figure \ref{fig-a1b}.
As the dependence of \(a_1\) on the threshold is very weak
(the only $s_0^K$-dependent  contributions are suppressed by \(m_s^2/M^2\) or \(\alpha_s\)), we use the value of \(s_0^K\) obtained from the
sum rule for \(f_K\). Our confidence in the
sum rule is also supported by the fact that the result for $a_1^K$
is quite stable with respect to $M$ within the Borel window
(see figure \ref{fig-a1a}).
Finally we obtain for $a_1^K$ the interval (\ref{a1number}). 
The estimated spread of our prediction is obtained by varying all
input parameters within their allowed intervals 
(with \(0.8~\mbox{GeV}^2<M^2<1.5~\mbox{GeV}^2\)) and summing up the
resulting variations of $a_1^K$ in quadratures. 
If we tried to reach certain cancellations
and estimate $a_1^K$ dividing (\ref{eq-a1sumrule}) by the sum rule for $f_K^2$
(and not using the experimental value of $f_K$)
the result would be practically the same.

Obtaining the interval (\ref{a1number}), we tacitly assume the absence
of large SU(3)-violating contributions originating from 
the higher-order terms of OPE (with dimension larger than 6) and not 
included in the diagonal sum rule. 
In fact, within the standard QCD sum rule approach there are two major sources 
of SU(3)-violation in OPE: the strange quark mass and the strange/nonstrange 
quark condensate ratio, all other effects, being derivatives of these two
(assuming factorization of higher-dimensional condensates), are under control. 
An argument in favour of this conjecture is provided by the fact  
that one of the typical SU(3)-violating effects, the ratio
$f_K/f_\pi$ is successfully reproduced \cite{KMM} from the sum rule 
derived from the same diagonal correlation function with dimension-6 accuracy.

The fact that the continuum contribution to the sum rule (\ref{eq-a1sumrule})
is small, indicates a relative suppression of
hadronic states heavier than the kaon. The correlation function of axial-vector currents
receives contributions from both pseudoscalar and axial-vector hadronic states.
Importantly, the excited pseudoscalar resonances
(the candidates for the lowest states
\cite{PDG} are $K(1460)$ and $K(1830)$) have  decay constants
proportional to light-quark masses and suppressed
with respect to $f_K$. E.g., the sum rule estimates obtained in \cite{Malt}
yield $f_{K(1460),K(1830)} <20$ MeV. Therefore, if one adds
the contributions of excited $K$-resonances to the ground-state kaon term on l.h.s.
of (\ref{sumrule1}) (increasing the threshold $s_0^K$ correspondingly)
one gets relative suppression factors $(f_{K(1460),K(1830)}/f_K)^2<2\%$
allowing to neglect these additional contributions altogether.

To investigate the impact of the axial-vector states we  
attempted to include explicitly
the contribution from the lowest resonance  \(K_1(1270)\) \cite{PDG} in the sum rule, in addition to the ground-state $K$-meson contribution. The threshold parameter
$s_0^K$ is then correspondingly increased. We calculated 
the \(K_1\) contribution from an independent QCD sum rule based on a 
correlation function where only the axial-vector states contribute. 
The procedure is shown in more details in appendix \ref{App-K1}.
Putting the result for $K_1$ into the hadronic representation of (\ref{eq-corr})
shifts the central value for \(a_1^K\) upwards by no more than \(20\%\).
From that we conclude that: 1) quark-hadron duality works reasonably well
and 2) upon the inclusion of \(K_1\), \(a_1\) moves even further from
negative values.

As a byproduct, we also calculated the second Gegenbauer moment of the kaon DA
and obtain:
\be
a_2^K(1~ \mbox{GeV})=0.27_{-0.12}^{+0.37}\,,
\label{a2}
\ee
in agreement (within uncertainties) with the estimate of \cite{BallB,KMM}
obtained from the same diagonal sum rule.
Furthermore, putting \(m_s\to 0, \langle\bar{s}s\rangle\to\langle\bar{q}q\rangle\),
we also get an estimate for the second Gegenbauer moment of the pion:
\be
a_2^\pi(1~ \mbox{GeV})= 0.26_{-0.09}^{+0.21}\,,
\label{a2pi}
\ee
in the ballpark of other current estimates of this parameter
(see, e.g \cite{BK,BMS}).
Finally, we have checked that the moments (\ref{eq-sumrule}) become unstable
for \(n>2\) which is expected.

\section{Is the nondiagonal sum rule reliable ?}

Our new result for \(a_1^K (1 ~\mbox{GeV})\) significantly differs from
the estimate (\ref{BBnumber}) obtained in \cite{BallB} from the nondiagonal sum rule.
The deviation remains
even if we stretch both numbers towards each other by adding/subtracting
the
estimated uncertainties. The difference is
also qualitative, because we predict a positive sign for the
asymmetry $\langle x_s-x_{u,d}\rangle_K $. In fact, our $a_1^K$
has the same sign as (\ref{CZa1})
obtained in \cite{CZ}
from the diagonal sum rule \footnote{
The value obtained in \cite{CZ} from the nondiagonal sum rule is plagued by the error
traced in \cite{BallB} and should be ignored.}.
The  fact that our estimate is smaller than (\ref{CZa1}), reflects the importance
of the corrections taken into account in our calculation
and absent in \cite{CZ}.

\begin{figure}
  \centering
  \subfigure[]{
    \centering\shortstack{\includegraphics[width=0.46\textwidth]{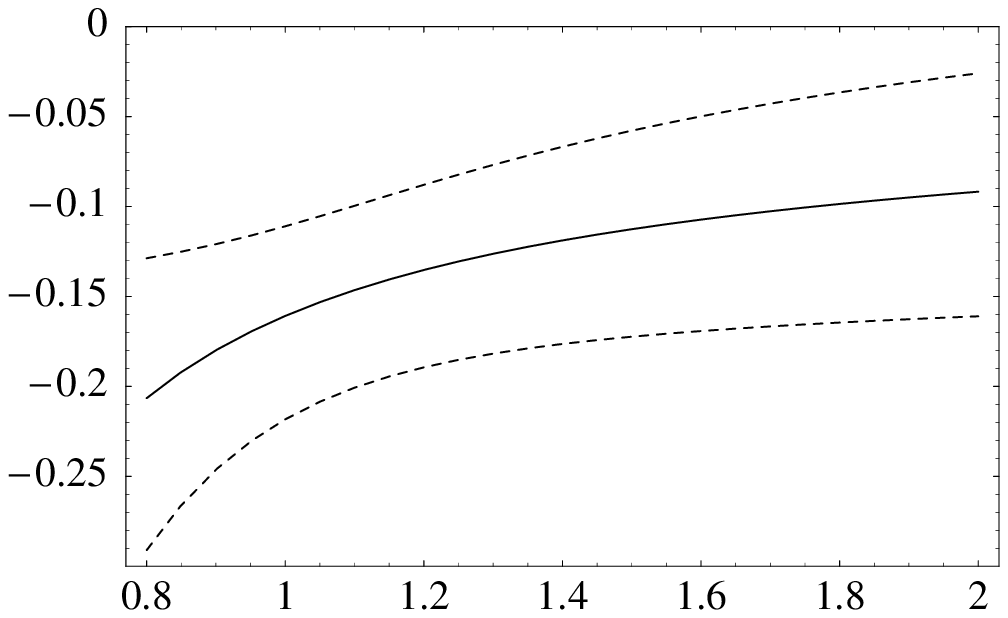}\\
             \small\(M^2/GeV^2\)} \label{fig-nondiag_a}
    }%
  \subfigure[]{
    \centering\shortstack{\includegraphics[width=0.46\textwidth]{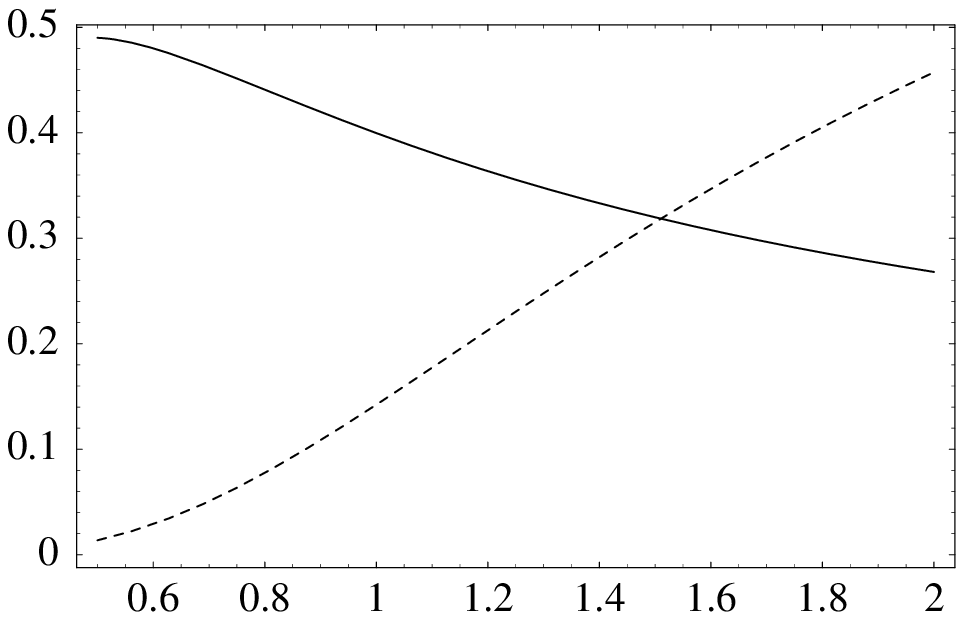}\\
             \small\(M^2/GeV^2\)} \label{fig-nondiag_b}
    }%
  \caption{{\it a) \(a_1^K\) from the nondiagonal sum rule obtained in \cite{BallB} (solid) with estimated uncertainties (dashed);
b) the dimension-six contribution (solid) and continuum (dashed) contributions
divided by the r.h.s. of the nondiagonal sum rule. }}
\end{figure}

Let us take a closer look at the nondiagonal sum rule. Numerically
evaluating the expressions presented in \cite{BallB}, we plot
the result for $a_1^K$ as a function of $M^2$
in Fig.~\ref{fig-nondiag_a} where we have used
the same input and estimated the uncertainties in the same way
as for the sum rule (\ref{eq-a1sumrule}).
First of all, one notices that in this sum rule
the three  numerically large contributions: 1) the $O(\alpha_s^0)$
perturbative term, 2) the tree-level quark-condensate and 3)
the quark-gluon condensate terms, almost cancel each other and this
cancellation seems to happen accidentally. The remaining large contributions
are the first-order in $\alpha_s$ terms in the perturbative and
quark-condensate parts. Comparing $O(\alpha_s)$ and $O(\alpha_s^0)$
terms separately, we see that  the former is about two times larger
than the latter in the perturbative part\footnote{%
We have found that also for
\(n=0\), the \(O(\alpha_s)\) term in the nondiagonal sum rule
is larger than \(O(\alpha_s^0)\) one.}.
Both effects: the cancellation of the leading terms and
enhancement of subleading terms cast doubt on
the accuracy of the sum rule and on the validity of the perturbative expansion.
Moreover, since the nondiagonal correlation function
contains a pseudoscalar current, one has to worry
about potentially important instanton effects, not taken into
account in the local OPE series. This issue, however, deserves a separate study.

Assessing the nondiagonal sum rule further, we notice that
a Borel window for it practically does not exist.
Indeed,
as can be seen from Fig.~\ref{fig-nondiag_b}, there is no region
of \(M^2\), in which both the convergence of the OPE, manifested by the smallness
of the dimension six contribution (the term with \(m_s\) multiplied
by the quark-gluon condensate), and the
suppression of excited states are guaranteed. In particular, \(a_1^K\)
is very sensitive to the choice of the threshold
parameter. This may signal a greater role of higher states. On the other hand,
only excited pseudoscalar $K$ resonances contribute to this sum rule, and as
already noticed, their terms are
suppressed with respect to the kaon ground-state term by the squares
of the small decay constants. The situation is therefore somewhat controversial.
In \cite{BallB}
the reliability of the nondiagonal sum rule is
supported by obtaining $a_0^K=1.05- 1.1$, close to 1.
In fact, we repeated the numerical analysis of the nondiagonal sum rule for \(a_0\)
with our input parameters, and could not find
a reasonable value for \(s_0^K\) for which the Borel stability takes place.
The value \(s_0^K=1.8~\mbox{GeV}^2\) given in \cite{BallB} seems to be somehow arbitrary.
We conclude that the nondiagonal sum rule suffers from problems
which are not yet fixed and therefore this sum rule is 
not reliable in its present form.

\section{ Constraining \(a_1^K\) with light-cone sum rules
 for the $K\to \pi$ form factor}

The form factors of pseudoscalar mesons at large momentum transfers
are among the most important hadronic observables calculated using DA's.
Here we concentrate on the $K\to \pi$ transition
form factor $f_{K\pi}^+$ defined
via hadronic matrix element
\be
\langle \pi^-(p-q)|\bar{s}\gamma_\mu u|K^0(p)\rangle =
2f_{K\pi}^+(q^2)p_\mu -(f_{K\pi}^+(q^2)-f^-_{K\pi}(q^2))q_\mu\,.
\label{matrKpi}
\ee
The form factor $f^+_{K\pi}$ is measurable at small timelike $0<q^2<(m_K-m_\pi)^2$ in $K_{l3}$ decays.
However, it is completely legitimate to consider $f^+_{K\pi}$ also at
large spacelike $q^2=-Q^2$ where it can be related to DA's of pion and kaon.
One well-known approach to calculate this form factor at large $Q^2$ is provided
by the method of light-cone sum rules (LCSR) \cite{lcsr}. The calculation
essentially repeats the applications of LCSR to the 
pion and kaon electromagnetic
form factors \cite{pionFF,BK}.
In particular, in \cite{BK} it was suggested to constrain 
the kaon DA by using two different LCSR
for the same form factor $f^+_{K\pi}(Q^2)$.
An illustrative calculation done in \cite{BK}
at the twist 2 level, showed that at $a_1^K>0$  
two different LCSR agree \footnote{ To avoid confusion, 
we note that in \cite{BK}
an opposite sign convention for $a_1^K$ is adopted.}.
Here we repeat that analysis with a greater accuracy, 
taking into account also higher-twist effects in LCSR.

One starts from a generic correlation function
\begin{equation}
T_{\mu\nu}=i\int d^4x e^{iqx}\langle 0|T\left\{(\bar{q}_2(0)\gamma_\mu\gamma_5q_1(0)),
(e_1\bar{q}_1(x)\gamma_\nu q_1'(x)+e_2\bar{q}_2'(x)\gamma_\nu q_2(x))\right\}|P(p)\rangle\,,
\label{corrLC}
\end{equation}
where $q_{1,2}$ are the light-quark fields and $P=\pi$ or $K$.
In both cases the correlation function
is expanded at large $Q^2$ near light-cone $x^2\sim 0$ up to twist-4 accuracy.

The first LCSR is then obtained from the invariant amplitude
multiplying the \(p_\mu p_\nu\) kinematical structure of the above correlation function
by putting \(q_1=s,\ q_1'=u,\ q_2=d,\ P=\pi^+,\ e_1=1,\ e_2=0\).
The result involves only the pion DA's starting from \(\varphi_\pi(u)\) 
and is naturally independent of \(a_1^K\).
For simplicity, we give here only the leading-twist expression
for the resulting LCSR \cite{BK}:
\begin{equation}
   f^+_{K\pi}(Q^2)=\frac{f_\pi}{f_K}\int_{u_0^K}^1 du~\varphi_\pi(u)
                         e^{-\frac{\bar{u}Q^2}{u M^2}+\frac{m_K^2}{M^2}}\,,
  \label{fKpi-from-phipi}
\end{equation}
where \(u_0^K=Q^2/(Q^2+s_0^K)\) and \(s_0^K\) is the duality threshold in the kaon channel, already determined above. The details of derivation and expressions
for the higher twists can be found in \cite{BK}.

The second LCSR  comes from a different flavour pattern in (\ref{corrLC}): \(q_1=d,\ q_2=u,\ q_2'=s,\ e_1=0,\ e_2=1,\ P=K^0\). In this case the correlation function
(\ref{corrLC}) reduces to a vacuum-to-kaon matrix element involving
(after light-cone expansion) the kaon DA's and containing \(a_1^K\).
Again, we only  show the leading twist-two piece of the second LCSR:
\begin{equation}
  f^+_{K\pi}(Q^2)=\frac{f_K}{f_\pi}\int_{u_0^\pi}^1 du~\varphi_K(u)
                         e^{-\frac{\bar{u}Q^2}{u M^2}-\frac{\bar{u}m_K^2}{M^2}}\,,
\label{fKpi-from-phiK}
\end{equation}
where \(u_0^\pi\) is related to the duality threshold \(s_0^\pi\)
in the pion channel by \(s_0^\pi=(1-u_0^\pi)(Q^2/u_0^\pi+m_K^2)\).

To evaluate the sum rules (\ref{fKpi-from-phipi}) and (\ref{fKpi-from-phiK})
numerically, we use the same input (and uncertainties) for the relevant DA parameters as in \cite{BK} (see  also \cite{KMM}),
but leave \(a_1^K\) as a free parameter.
At \(1~\mbox{GeV}^2<Q^2<3~\mbox{GeV}^2\), 
one and the same observable  $f^+_{K\pi}(Q^2)$ is then calculated 
from LCSR in two
different ways. This allows to constrain  \(a_1^K\) 
completely independent of the sum rule calculation done in previous sections.

\begin{figure}
  \centering
  \subfigure{\centering\shortstack{\includegraphics[width=0.3\textwidth]{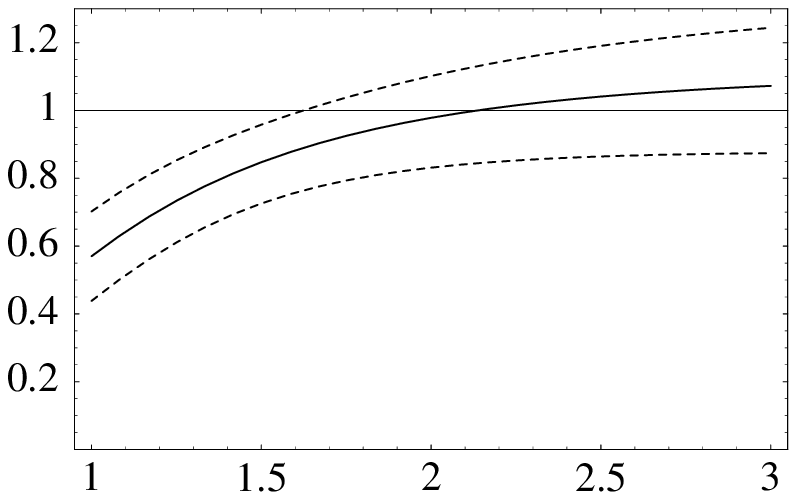}\\
             \small\(Q^2/GeV^2\)}\label{fig-fkpi-a1-pos}}
  \subfigure{\centering\shortstack{\includegraphics[width=0.3\textwidth]{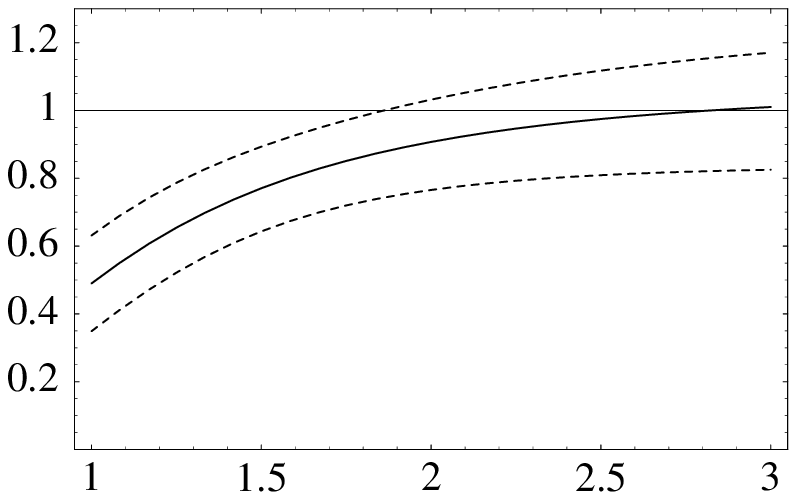}\\
             \small\(Q^2/GeV^2\)}\label{fig-fkpi-a1-0}}
  \subfigure{\centering\shortstack{\includegraphics[width=0.3\textwidth]{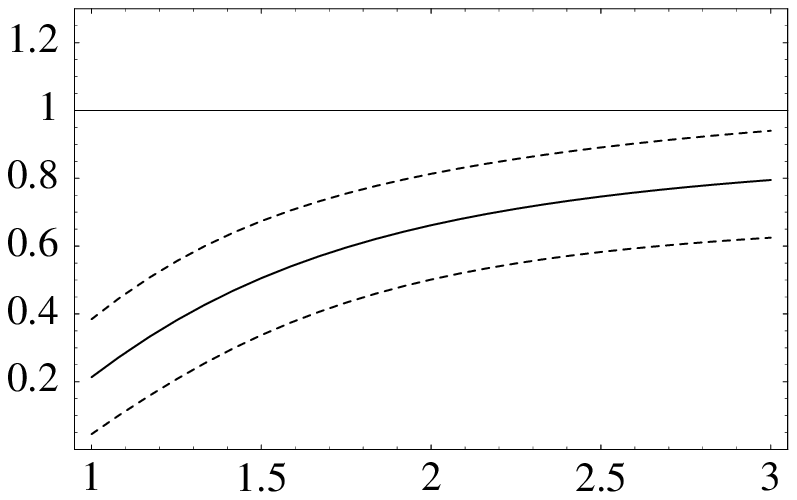}\\
             \small\(Q^2/GeV^2\)}\label{fig-fkpi-a1-neg}}
  \caption{{\it The ratio of two LCSR for $f^+_{K\pi}$  
defined as $r.h.s.(\ref{fKpi-from-phiK})/r.h.s.(\ref{fKpi-from-phipi})$
at $a_1^K= +0.05^{+0.02}_{-0.01}$ (left), $a_1^K=0$ (central) and 
\(a_1^K=-0.18\pm 0.09\)\cite{BallB} (right).
The dashed lines indicate the uncertainties from varying all
input parameters in the allowed ranges.}}
  \label{fig-fkpi-ratios}
\end{figure}

We calculate the ratio of r.h.s. of two sum rules (\ref{fKpi-from-phipi}) and
(\ref{fKpi-from-phiK}) as a function of \(Q^2\). (Note that in this ratio some
of uncertainties partially cancel).  Ideally, the ratio should be equal to one.
It is plotted in Fig.~\ref{fig-fkpi-ratios} for different values of \(a_1^K\).
The variation of the Borel parameter from \(1\) to \(2~\mbox{GeV}^2\) is included
within the uncertainties. We find that for a positive value of \(a_1^K\),
the ratio is indeed consistent with 1. Remarkably,
at \(a_1^K<0\) it noticeably deviates from 1.
Alternatively, one demands that the ratio of two LCSR is equal to 1 at, say,
the middle value \(Q^2=2~\mbox{GeV}^2\) and then solves it for \(a_1^K\),
yielding \(a_1^K\approx 0.2\pm 0.1\), 
which is consistent with (\ref{a1number}).

\section{SU(3)-Violation in \(B\) decays: an update}

Having the new estimate (\ref{a1number}) for the Gegenbauer moment $a_{1}^K$,
we find it appropriate to give an update of SU(3)-violating
effects in $B\to P$ form factors and $B\to PP$ decay amplitudes ($P=\pi,K$) 
calculated in \cite{KMM} from LCSR, where we have used the interval (\ref{BBnumber}) of $a_1^K$ from \cite{BallB}.  For consistency, in order to have correlated 
uncertainties,  we also use the estimates (\ref{a2}) and (\ref{a2pi})
for $a_2^K$ and $a_2^\pi$. All other input parameters in LCSR are
not changed. 

Note that in the SU(3)-violating parts of LCSR, \(a_1^K\) plays quite an
important role. To demonstrate that, below we display the
numerical predictions of LCSR (for the central values of parameters
adopted in \cite{KMM}) isolating the parts proportional
to $a_1^K$: 
\begin{eqnarray}
  f_{B\to K}^+(0)&=&(1.31+1.11\cdot a_1^K(\mu_B))f_{B\to\pi}^+(0)\,, \nonumber\\
  f_{B_s\to K}^+(0)&=&(1.25-1.02\cdot a_1^K(\mu_B))f_{B\to\pi}^+(0)\,.
\label{relFF}
\end{eqnarray}
In the above, the rest of SU(3)-violation is caused by differences
between other parameters in the kaon and pion channel, mainly
between $f_K$ and $f_\pi$,
$s_0^K$ and $s_0^\pi$, $a_2^K$ and $a_2^\pi$.
The SU(3)-violation in the parameters of higher-twist DA's
produce minor effects.
Note that in (\ref{relFF}) the characteristic scale of
$\mu_B\sim O(m_B^2-m_b^2)$ is used, and the scale-dependent parameters,
such as $a_1^K$ itself, have to be evolved to this scale
using the known anomalous dimensions.

In obtaining the updates given below we follow the same 
procedure of estimating  uncertainties as in \cite{KMM}, that is,
the analytic expressions are used and all input parameters are varied 
in a correlated way.

First, we present the updated values of the \(B\to\pi, K\) form factors
which simply follow from (\ref{relFF}) substituting our new estimate
of $a_1^K$:
\begin{eqnarray}
  \frac{f^+_{BK}(0)}{f^+_{B\pi}(0)}
    = 1.36^{+0.12}_{-0.09}\,, ~~
  \frac{f^+_{B_s K}(0)}{f^+_{B\pi}(0)}
    = 1.21^{+0.14}_{-0.11}\,.
\end{eqnarray}
Note that these ratios differ mainly by the sign of the \(a_1^K\)
contribution and are therefore interchanged in their numerical value
with respect to the ratios given in \cite{KMM} after a sign change
of \(a_1^K\). Consequently, the pattern of \(SU(3)\)-violation
in the factorizable part of the amplitudes also changes considerably:
\begin{equation}
\left.
\begin{array}{r@{\quad}lc}
A_{fact}(B\to \pi K)&=&\frac{f_K}{f_\pi}A_{fact}(B\to\pi\pi)=1.22\\
&&\\
A_{fact}(B\to K \pi)&=&1.36^{+0.12}_{-0.09}\\
&&\\
A_{fact}(B\to K \bar{K})&=&1.65^{+0.14}_{-0.11}\\
&&\\
A_{fact}(B_s\to K\bar{K})&=&1.52^{+0.18}_{-0.14}\\
&&\\
A_{fact}(B_s\to K \pi)&=&1.25 ^{+0.14}_{-0.12}\\
\end{array} \right\}\times A_{fact}(B\to\pi\pi)\,.
\label{ratios}
\end{equation}
In our notation of the factorizable amplitudes above
the first (second) meson in the final state is the
one containing the spectator quark of $B$ meson (the ``emitted'' one).
Interestingly, the violation of the \(SU(3)\)-relation \cite{GLR}
\begin{equation}
A(B^-\to \pi^-\bar{K}^0) +\sqrt{2}A(B^-\to \pi^0 K^-)
=\sqrt{2}\left(\frac{V_{us}}{V_{ud}}\right) A(B^-\to \pi^-\pi^0)
\{1+\delta_{SU(3)}\}\,,
\label{rel1}
\end{equation}
determined by the parameter $\delta_{SU(3)}$
is robust with respect to $a_1^K$, and the prediction obtained in
\cite{KMM} (including nonfactorizable effects)
remains nearly unchanged:
\begin{equation}
\delta_{SU(3)}=(0.215^{+0.019}_{-0.016})+(-0.009^{+0.009}_{-0.010})i.
\end{equation}
Further improvements in the input parameters in LCSR, in particular,
a more precise determination of Gegenbauer moments will allow to decrease
the uncertainties in (\ref{ratios}) and (\ref{rel1}).

\section{Conclusions}

In this paper we have reanalyzed the QCD sum rule prediction
for the first Gegenbauer moment $a_1^K$ in the twist-2 DA of the kaon.
This is an important SU(3)-violation effect reflecting the momentum asymmetry
of strange and nonstrange quarks in the kaon. The advantage of the sum
rule method is the ability to connect the value of $a_1^K$ directly to the
fundamental QCD parameters: the mass of the $s$ quark
and the difference between $s$- and $u,d$- quark condensates.

The diagonal correlation function provides
a new estimate of  $a_1^K$ which has the same positive sign 
as in the original calculation
of \cite{CZ} but a smaller value
after including important perturbative correction
to the quark condensate. We investigated the quality of this sum rule and found
that all usual criteria are satisfied: Borel stability, hierarchy of
power corrections, smallness of the higher-state contributions.

The OPE for the nondiagonal correlation
function has been significantly improved in \cite{BallB}.
However, taking into account the numerical analysis presented
in sect. 4, we think that the corresponding sum rule is numerically
not safe for $a_1^K$ extraction.
We therefore conclude that one should rely on the results from
the diagonal sum rule.

Our estimate of $a_1^K$ has an uncertainty of $\sim 30\%$. It can be slightly
improved further, e.g. calculating the $O(\alpha_s)$ correction to the
(chirally-suppressed) perturbative part of the diagonal sum rule. However,
the overall uncertainty will hardly become lower
than 15-20\%, due to the limited accuracy of the sum rule determination. 
Therefore,  it would be very interesting
to calculate $a_1^K$, that is, the matrix element (\ref{eq-matrel})  
from lattice QCD.
\bigskip

{\it Note added}

After this paper was completed, the work \cite{BraunLenz}
appeared, where a similar result for $a_1^K$ is obtained using
a different method of operator identities.
\bigskip

{\bf Acknowledgements}

We are grateful to V.~Braun and M.~Shifman for useful
discussions. This work is supported by  the DFG Sonderforschungsbereich
SFB/TR(``Computational Particle Physics'') and by the German Ministry for
Education and Research (BMBF).

\begin{appendix}
\section{$\alpha_s\langle\bar{q}q\rangle$ for $n=0,2,3$}
\label{App-Condensate-2}

In addition to (\ref{alphasqq}) we present here 
the \(n=0\) and \(n= 2,3\) projections of
the \(O(\alpha_s\langle\bar{q}q\rangle)\) contributions
to the sum rule (\ref{eq-sumrule}):
\begin{equation}
 \int_0^1 du C_0^{3/2}(u-\bar{u}) \pi^{\alpha_s\langle\bar{q}q\rangle}(u,q^2)=
 -\frac{3}{4}\frac{\alpha_s C_F}{\pi}\frac{m_s\langle\bar{s}s\rangle}{(q^2)^2}
-\frac{\alpha_s C_F}{\pi}\frac{m_s\langle\bar{u}u\rangle}{(q^2)^2}
\end{equation}

\begin{align}
 \int_0^1 du C_2^{3/2}(u-\bar{u}) \pi^{\alpha_s\langle\bar{q}q\rangle}(u,q^2)&=
  \frac{\alpha_s C_F}{\pi}\frac{m_s\langle\bar{s}s\rangle}{(q^2)^2}
  \left[-\frac{25}{4}\left(\Delta-\log\left(-\frac{q^2}{\mu^2}\right)\right)-\frac{763}{24}\right]
\nonumber
\\
&-\frac{\alpha_sC_F}{\pi}\frac{7}{2}\frac{m_s\langle\bar{u}u\rangle}{(q^2)^2},
\end{align}

\begin{align}
 \int_0^1 du C_3^{3/2}(u-\bar{u}) \pi^{\alpha_s\langle\bar{q}q\rangle}(u,q^2)&=
  \frac{\alpha_s C_F}{\pi}\frac{m_s\langle\bar{s}s\rangle}{(q^2)^2}
  \left[-\frac{157}{12}\left(\Delta-\log\left(-\frac{q^2}{\mu^2}\right)\right)+\frac{24629}{360}\right]
\nonumber
\\
&-\frac{\alpha_sC_F}{\pi}\frac{9}{2}\frac{m_s\langle\bar{u}u\rangle}{(q^2)^2},
\end{align}

\section{Including axial-vector mesons in the sum rule}
\label{App-K1}
In order to estimate the contribution from the axial-vector meson 
to the diagonal sum rule for \(a_1^K\), we need a slightly modified 
correlation function as compared to (\ref{eq-corr}):
\begin{equation}
  \Pi_{\mu\nu}(q,z)=i\int d^4x e^{i q\cdot x}\langle 0|T\left\{\bar{u}(x)\gamma_\mu\gamma_5 s(x),
  \bar{s}(0)\gamma_\nu\gamma_5\left[0,z\right]u(z)\right\}|0\rangle,
\label{eq-corr-K1}
\end{equation}
which can be decomposed in five invariant amplitudes:
\begin{equation}
  \Pi_{\mu\nu}=\Pi_1 q_\mu q_\nu+\Pi_2 g_{\mu\nu}+\Pi_3 q_\mu z_\nu+\Pi_4 z_\mu q_\nu+\Pi_5 z_\mu z_\nu\,.
\end{equation}
Note that the amplitude \(\Pi_1\) above  corresponds to 
\(\Pi\) from (\ref{eq-corr}):  
\((q\cdot z)^2 \Pi_1(q^2,q\cdot z)=\Pi(q^2,q\cdot z)\).
For simplicity, we only take into account the lowest \(K_1(1270)\) state.
The second  axial-vector resonance \(K_1(1400)\) \cite{PDG} 
can also be included provided one adjusts the continuum threshold properly.
But that will not noticeably change the effect of axial-vector states 
on the kaon contribution in the sum rule.

We proceed by defining the relevant hadronic matrix elements:
\begin{eqnarray}
  \langle 0|\bar{u}\gamma_\mu\gamma_5 s|K_1(q,\lambda)\rangle
  &=&\epsilon_\mu^\lambda f_{K_1} m_{K_1}\,,\\
  \langle K_1(q,\lambda)|\bar{s}(0)\gamma_\nu\gamma_5 [0,z] u(z)|0\rangle
  &=&\epsilon_\nu^{\lambda *} f_{K_1}m_{K_1} \int_0^1 du~ e^{i\bar{u} q\cdot z}\varphi_{K_1}(u)\,,
\end{eqnarray}
where $\varphi_{K_1}(u)$ is one of the DA's of $K_1(1270)$ and
$\epsilon$ is its polarization vector.
Using the above definitions,  one writes down the \(K_1\) 
contribution to (\ref{eq-corr-K1}):
\begin{equation}
  \Pi^{K_1}_{\mu\nu}(q,z)=\left(q_\mu q_\nu-g_{\mu\nu} m_{K_1}^2\right)
\frac{ f_{K_1}^2}{m_{K_1}^2-q^2}\int_0^1 du~ e^{i\bar{u} q\cdot z}\varphi_{K_1}(u).
\end{equation}
The kaon term is present only in \(\Pi_1\), whereas
\(K_1\) contributes to both \(\Pi_1\) and \(\Pi_2\),
and due to transversality, $\epsilon^\lambda\cdot q=0$,
these two contributions are equal. Therefore, we can 
estimate the $K_1$ contribution  from the sum rule
for the invariant amplitude $\Pi_2$:
\begin{equation}
-m_{K_1}^2f_{K_1}^2 a_n^{K_1} e^{-m_{K_1}^2/M^2}=\frac{2(2n+3)}{3(n+1)(n+2)}\frac{1}{\pi}\int_0^{s_0^{K_1}}ds~e^{-s/M^2}\int_0^1 du~C_n^{3/2}(u-\bar{u})\mbox{Im}_s\pi_2(u,s)\,.
\label{sumrule2}
\end{equation}
and use the estimate of  $f_{K_1}^2 a_1^{K_1}$ in the sum rule 
for $\Pi_1$.
Since a rough estimate is sufficient for our purposes, we only include
in $\pi_2$ the most important contributions of
the perturbative loop, quark condensate and quark-gluon condensate.
We then turn  to the sum rule
for \(\Pi_1=\int_0^1 du~e^{i\bar{u}qz}\pi_1(u,q^2)\),
where both the kaon and \(K_1\) contribution are now included. After projecting
out the \(n^{th}\) moment, we get:
\begin{eqnarray}
&&f_K^2 a_n^K e^{-m_K^2/M^2}+f_{K_1}^2a_n^{K_1} e^{-m_{K_1}^2/M^2}\nonumber\\
&&=\frac{2(2n+3)}{3(n+1)(n+2)}\frac{1}{\pi}\int_0^{s_0^{K_1}}ds~e^{-s/M^2}\int_0^1 du~C_n^{3/2}(u-\bar{u})\mbox{Im}_s\pi_1(u,s).
\label{srK1}
\end{eqnarray}
Here \(s_0^{K_1}\) is the new threshold, which naturally 
lies above \(m_{K_1}^2\). Note that \(\pi_1(u,q^2)\) was already calculated 
for the sum rule (\ref{eq-a1sumrule}).

Although the sum rules for \(\Pi_1\) and \(\Pi_2\) are different, 
we observe the best  Borel stability of both by using the same duality 
threshold \(s_0^{K_1}\).
As before, we first put \(n=0\) in (\ref{sumrule2})
and choose \(s_0^{K_1}\) so that maximal
Borel stability for \(f_{K_1}^2\) is achieved. We get
\(f_{K_1}^2\approx {0.031^{+0.006}_{-0.003}~\mbox{GeV}^2}\)
(in a good agreement with the experimental value of this decay constant
extracted from the $\tau\to K_1(1270)\nu_\tau$ partial width)
and \(s_0^{K_1}\approx 1.7~\mbox{GeV}^2\).
Furthermore, switching to \(n=1\) in (\ref{sumrule2})
and dividing by \(f_{K_1}^2\),
we then  obtain \(a_1^{K_1}\approx -0.04^{+0.04}_{-0.03}\).
Finally, our ``$K_1$-improved'' value of \(a_1^K\)
obtained from (\ref{srK1}) is
\begin{equation}
a_1^K=0.07^{+0.02}_{-0.03}\,,
\end{equation}
slightly above the original one in (\ref{a1number}).

\end{appendix}

\end{document}